# Ferromagnetic Quantum Critical Point in FeGa$_3$ Tuned by Mixed-Valence Fe Dimers

N. Haldolaarachchige[1], J. Prestigiacomo[1], W. Adam Phelan[2], Y. M. Xiong[1], Greg McCandless[2], Julia Y. Chan[2], J. F. DiTusa[1], I. Vekhter[1], S. Stadler[1], D.E. Sheehy[1], P.W. Adams[1] and D.P. Young[1*]

[1]Department of Physics and Astronomy, Louisiana State University, Baton Rouge, Louisiana 70803, USA
[2]Department of Chemistry, Louisiana State University, Baton Rouge, Louisiana 70803, USA

*dyoung@phys.lsu.edu

The magnetic, transport, and thermal properties of single crystals of the series Fe(Ga$_{1-x}$Ge$_x$)$_3$ are reported. Pure FeGa$_3$ is a nonmagnetic semiconductor, that when doped with small concentrations of Ge (extrinsic electrons), passes through an insulator-to-metal transition and displays non-Fermi liquid (NFL) behavior. Moreover, we observed clear signatures of a ferromagnetic quantum critical point (FM-QCP) in this system at $x$ = 0.052. The mechanism of the local moment formation is consistent with a one-electron reduction of Fe dimer singlets – a unique structural feature in FeGa$_3$ – where the density of these mixed valence [Fe(III)-Fe(II)]$^2$ dimers provides a unique tuning parameter of quantum criticality.

## I. INTRODUCTION

FeGa$_3$ is a nonmagnetic intermetallic semiconductor that displays a large, negative thermopower near room temperature.[1, 3] Recent studies have shown that this system responds well to chemical doping, and its thermoelectric figure of merit has been increased, in some cases by a factor of five.[3] From a more holistic viewpoint, the nature of the band gap in transition metal semiconductors has spurred comparisons between FeGa$_3$ and other semiconductors, such as FeSi and FeSb$_2$, which have shown evidence of strong correlation effects.[4-7]

Recent photoemission measurements[8] on single crystals of FeGa$_3$ indicated a valence band maximum (VBM) composed primarily of a Ga 4$sp$ band with the Fe 3$d$ bands located at lower energy. The energy gap was found to be ~0.4 eV, in good agreement with transport measurements, and they could model the energy band dispersions with an (LDA + $U$) calculation, where $U_{\text{eff}}$ ~ 3 eV. The additional on-site Coulomb repulsion results in a decrease in the energy separation of the two Fe 3$d$-derived bands and a decrease in their dispersion widths.[8] However, their ARPES measurements of the electronic states near the VBM showed no temperature dependence, which is in contrast to FeSi, where a sharp peak at the VBM grows in intensity with decreasing temperature.[9] Furthermore, there is a large Fe-3$d$ DOS near the gap edge in FeSi, with the gap being ~7 times smaller than it is in FeGa$_3$. Strong correlation effects in FeSi appear when charge carriers are thermally excited across this small, renormalized gap. Such activation of carriers in FeGa$_3$ doesn't occur until much higher temperature (~500 K). The larger gap, and the location of the Fe 3$d$ bands at lower energy relative to the band gap edge, provides an explanation why electron correlations in pure FeGa$_3$ appear weak compared to FeSi.

However, Yin and Pickett have recently noted[10] that the four Fe atoms per unit cell in FeGa$_3$ exist structurally as two Fe-Fe dimers, where the Fe-atom separation distance within the dimer is 2.77 Å, compared to 2.48 Å in bcc Fe metal. Having Fe atoms anti-aligned in each dimer (a band singlet) would offer an alternative explanation for the non-magnetic ground state of FeGa$_3$. They further suggest that FeGa$_3$ is near a magnetic instability, and their first-principles LDA calculations predict an ordered magnetic state when including an on-site Coulomb repulsion term ($U \sim 2$ eV).[10] Previous work on FeGa$_3$ by us[3] and other groups[11] has, in fact, shown that the electronic structure and physical properties, such as thermoelectric efficiency, are extremely sensitive to chemical doping, even at small doping levels (< 5%). Bittar $et$ $al$.[11] showed that cobalt doping leads to weakly-coupled local moment formation with a crossover to a metallic state and strongly correlated electron behavior similar to that observed in heavy Fermion systems. Similar gap suppression by chemical doping is well documented in both FeSi and FeSb$_2$.[12-15] Our current doping studies were motivated by these earlier results and the prospect of novel magnetic ground states. While NFL behavior and a FM instability in Fe(Ga$_{1-x}$Ge$_x$)$_3$ were independently observed by Umeo $et$ $al$.[16], we show below clear evidence of the FM-QCP in the magnetic, transport, and thermal properties of this system and propose a model that captures the salient features of the data – one that is intimately tied to the material's Fe-dimer structure.

## II.    EXPERIMENTAL

Single crystals of pure FeGa$_3$ and the Ge-doped samples were synthesized by a standard metallic flux technique[17] using excess Ga metal. All of the starting materials were purchased from Alfa Aesar and had a purity of at least 99.999%. A molar ratio of 1:5(1–$x$):5$x$ [Fe:Ga:Ge][1,18] was used, assuming a stoichiometric formula of Fe(Ga$_{1-x}$Ge$_x$)$_5$ in order to obtain the nominal concentrations of Ge. The crystal structure and phase purity of all the samples were investigated by single crystal and powder X-ray diffraction. A small crystal fragment was glued to a glass fiber and mounted on the goniometer of a Nonius Kappa CCD diffractometer equipped with Mo-$K_\alpha$ radiation ($\lambda = 0.71073$ Å) and data were collected at 290 K. Elemental analysis was performed using wavelength dispersive X-ray spectroscopy (WDS) with a JEOL JXA-733 SuperProbe Electron Probe Microanalyzer (EPMA). Microprobe analysis confirmed that the doping percentages closely matched the nominal concentrations. Electrical resistivity was measured using a standard four-probe method in a Quantum Design Physical Property Measurement System (PPMS) using a bar-shaped sample (1 mm × 1 mm × 2 mm) from 300 K to 1.8 K. The specific heat was measured in the PPMS using a time-relaxation method between 0.4 and 20 K. The temperature dependence of the magnetic susceptibility and magnetization versus applied field at 3 K were measured either with the PPMS or a Quantum Design squid magnetometer (MPMS).

## III. RESULTS AND DISCUSSION

Our initial effort focused on the series Fe(Ga$_{1-x}$Ge$_x$)$_3$. The tetragonal crystal structure of pure FeGa$_3$ ($P4_2/mnm$) is shown in Fig. 1a, where Fe-Fe dimer pairs (blue atoms) exist along the (110) direction in the $z = 0$ plane and along the $(1\bar{1}0)$ direction in the $z = ½$ plane. There are two unique Ga (Ge) sites: one of lower symmetry (pink spheres), forming a slightly corrugated net stacked along the $c$-axis, and a higher symmetry site (yellow spheres), located in the Fe-Fe dimer planes. Figures 1c and 1d show that the variation of lattice parameters as a function of $x$ (Ge doping), as determined by single crystal X-ray diffraction measurements, agrees well with Vegard's law. The calculated unit cell parameters ($a = 6.267$ Å and $c = 6.561$ Å) of pure FeGa$_3$ are in good agreement with previously reported data.[1, 19, 20] All the Ge-doped samples were free from detectable impurity phases or other elemental impurities. Results from wavelength dispersive X-ray (WDS) spectroscopy measurements (Fig. 1b) indicate the Ge concentration in the single crystal samples matches the nominal value.

Increasing the Ge concentration results in a shrinking of the unit cell along the $c$-axis (Fig. 1d and 1e), as expected by substituting the smaller radii Ge-atom for Ga. At the same time we observe that the $a$-axis lattice parameter is much less sensitive to doping and remains essentially unchanged (Fig. 1c), if not slightly increasing to within the resolution of the diffraction technique. One parameter of particular interest is the Fe-Fe distance in the dimers and its variation with Ge doping. For pure FeGa$_3$ we find the Fe-Fe distance (Fig. 1f) in the dimer to be ~ 2.77 Å which agrees well with published values.[5] The data indicate that the cell dimensions in the $ab$-plane are not largely affected by the small doping levels (≤10%) explored here. Thus, the Fe-Fe dimer distance remains fairly constant across the series.

The dc-magnetic susceptibility as a function of temperature is shown in Figs. 2a and 2b for single crystals of Fe(Ga$_{1-x}$Ge$_x$)$_3$. Pure FeGa$_3$ (not shown) was found to be diamagnetic below room temperature, in agreement with earlier reports.[1, 21] At a low doping level ($x$~ 0.01) the sample becomes paramagnetic (PM), and below ~200 K, the magnetic susceptibility of the doped samples could be well fit to a Curie-Weiss law, $\left[\chi(T) = \left(C/(T-\theta)\right) + \chi_0\right]$, where $C$ is the Curie constant, $\theta$ is the Weiss temperature, and $\chi_0$ is a temperature independent background term. No significant variations were observed between the zero-field-cooled (ZFC) and field-cooled (FC) curves.

For $x > 0.05$, the low-temperature susceptibility increases significantly (Fig. 2b), and an unexpected ferromagnetic (FM) state develops whose Curie temperature ($T_c$) increases with increasing $x$. The temperature derivative of the susceptibility, $d\chi/dT$, clearly shows the ordering temperature scaling with $x$ (Fig. 2c). The Curie temperatures of the FM samples were determined from Arrott plots[22] ($M^2$ vs $H/M$) (e.g. in Fig. 2d for the $x = 0.10$ sample). The isotherm which passes through the origin identifies the Curie temperature, $T_c$, and should be linear in $H/M$. The data in Fig. 2d show this isotherm corresponds to $T = 32$ K for $x = 0.1$ sample. However, the isotherm is not linear due to a surprising and considerable magnetic field dependence in $T_c$. Figure 2e demonstrates this dependence for the $x = 0.052$ sample. At $H = 0$, the sample is a $T_c = 0$ K ferromagnet, and $T_c$ increases with increasing field, following a power law dependence, where $T_c \sim H^n$, with $n \approx 0.15$.

The magnetization versus applied field ($M$ vs $H$) at 3 K (Fig. 2f) shows the requisite FM behavior with a saturated moment ($m_{sat}$) that, like $T_c$, increases with increasing $x$ (Fig. 3a). The high field magnetization value of the $x = 0.05$ sample is ~0.01 $\mu_B$/Fe, and this develops into a saturated moment $m_{sat} = 0.14$ $\mu_B$/Fe at $x = 0.10$. The effective magnetic moment ($\mu_{eff}$) can be calculated from the value of the Curie constants obtained from the fits to the data. Figure 3b shows the effective magnetic moment ($\mu_{eff}$) plotted as a function of $x$, both as $\mu_B$/Fe (lower curve) and $\mu_B$/Ge (upper curve), which also corresponds to $\mu_{eff}$/(doped electron). The effective moment per Fe atom (Fig. 3b) increases with $x$ from 0.18 $\mu_B$/Fe at $x = 0.01$ to 1.24 $\mu_B$/Fe at $x = 0.10$. A small increase in the slope of $m_{eff}$ versus $x$ is observed near $x \sim 0.05$. Given the values of $\mu_{eff}$ and $m_{sat}$, we can calculate a useful parameter in classifying the ferromagnetism – the Rhodes-Wolfarth ratio (RWR).[23] An RWR value near 1, such as for EuO and Gd, indicate localized magnetism, while larger values, i.e. those of ZrZn$_2$ and Co-doped FeSi, indicate an itinerant system. The value of RWR for Fe(Ga$_{1-x}$Ge$_x$)$_3$ is ~7 (Fig. 3c), suggesting the ferromagnetism is considerably itinerant. Also, it is interesting to note that the degree of localized moment in the series is essentially independent of $x$ for $x \leq 0.10$.

A conventional (classical) second-order magnetic phase transition is driven by thermal fluctuations and occurs at a finite temperature. In certain magnetic systems this phase transition can be suppressed to absolute zero temperature via a non-thermal tuning parameter, such as chemical doping, physical pressure, or magnetic field.[24-26] The magnetic data in Figs. 2a and 2b indicate the ferromagnetism develops for $x > $ ~0.05, suggesting the Curie temperature can be tuned to $T = 0$ K. Based on the magnetic data, we determined this critical concentration to be $x_c = 0.052 \pm 0.001$. The low temperature magnetic susceptibility $\chi(T)$ is plotted versus $T^{-4/3}$ in Fig. 3d for $x = 0.05, 0.052,$ and $0.06$. The susceptibility is quite sensitive to the doping level in this range due to the transition to a FM state. The data for $x = 0.052$ are linear in $T^{-4/3}$, indicating the magnetic susceptibility follows the power-law dependence predicted to occur near a FM QCP.[18, 27, 28]

The resistivity of pure FeGa$_3$ (Fig. 4a, solid circles, left axis) displays a complex temperature dependence with insulating behavior at low temperature. Above 260 K the data follow an activation law, between 260 K and 60 K, the resistivity decreases with cooling, and for temperatures below 60 K, the resistivity increases by almost five orders of magnitude. These different temperature regimes were studied previously in detail, and our results are in good agreement with the earlier work.[7] Upon Ge doping (Fig 4a, right axis), the resistivity becomes metallic over the entire measured temperature range, even for $x = 0.01$, with the room temperature resistivity dropping by over two orders of magnitude. For higher doping levels the metallic state remains, and the resistivity is further reduced. For samples with $x > 0.052$, i.e. FM, a small kink and corresponding decrease in the resistivity is observed at the Curie temperature, resulting from a reduction in the spin-disorder scattering (e.g. Fig. 4a inset for $x = 0.10$).

The low temperature resistivity for $x$ values close to $x_c$ was examined in greater detail by fitting the data to a power law[27]: $\rho(T) = \rho_0 + AT^n$, where $\rho_0$ is the residual resistivity at $T = 0$ K, $A$ is a generalized Fermi liquid (FL) coefficient, and $n$ the temperature exponent. The PM sample with $x = 0.04$ and the FM sample with $x = 0.06$ are well fit (Fig. 2b) by the above power law with $n = 2.0$, indicating a $T^2$ temperature dependence and behavior approximating a pure Fermi liquid (FL) and/or electron-magnon scattering in the ordered state.[29, 30] However, at $x = x_c$, the resistivity varies as $T^{5/3}$ ($n = 1.67$). This non-Fermi liquid (NFL) temperature dependence of the resistivity occurs only near the critical concentration at $x = 0.052$, and like the magnetic data, is characteristic of a FM QCP.[27, 31] Furthermore, the $A$ parameter (Fig. 4c) displays a sharp maximum over a narrow window near $x = x_c$.

For a normal FL, we expect the specific heat capacity to follow: $\frac{C}{T} = \gamma + \beta T^2$. In Fig. 5a we have plotted $\Delta C/T$, which represents the heat capacity with the $\beta T^2$ phonon term removed. The behavior of the specific heat capacity for $0.05 \leq x \leq 0.06$ is markedly sensitive to the Ge concentration (Fig. 5a). A significant enhancement in the Sommerfeld coefficient ($\gamma$) is observed with doping. The value of $\gamma$ for pure FeGa$_3$ was reported[7] to be 0.03 mJ/mol K$^2$, whereas for $x = 0.052$, $\gamma \sim 53$ mJ/mol K$^2$ as $T \rightarrow 0$ K. A similar mass enhancement was observed[11] in Co-doped FeGa$_3$.

At $x = x_c$, NFL behavior characterized by a logarithmic increase in the heat capacity toward zero temperature is observed. This is another clear indication of the system approaching a FM-QCP.[27, 28, 31] At higher concentrations, the logarithmic behavior is suppressed. The sample with $x = 0.09$ is FM ($T_c = 27$ K), and its low-temperature heat capacity tends toward saturation at just above 20 mJ/mol K$^2$ for temperatures below ~8 K. Figure 5b focuses on the low-temperature specific heat of the $x = 0.01$ sample (PM) and the $x = 0.09$ sample (FM). The low temperature behavior is clearly very different. In a ferromagnet, one expects magnons to contribute to the total specific heat with a $T^{3/2}$ temperature dependence. In Fig. 5b, we show that $\Delta C/T$ is linear in $T^{1/2}$ at low temperature for the FM sample, consistent with the magnon contribution.

The proposed phase diagram for Fe(Ga$_{1-x}$Ge$_x$)$_3$ is shown in Fig. 6, indicating the diamagnetic insulating (DI), paramagnetic (PM), and ferromagnetic (FM) phases as a function of $x$, as well as the FM-QCP that exists at $x = 0.052$. The inset of Fig. 6 shows that $T_c^{4/3}$ plotted versus $(x - x_c)$ is linear, which is the expected behavior near the FM-QCP.

Below we describe an empirical model which captures the salient features of the magnetic properties of the series. At the heart of the model is the unique Fe-Fe dimer that exists in the FeGa$_3$ structure. Materials containing transition metal dimers are currently under intense study for their possible use in magnetic data storage applications, as they represent the smallest bit size with magnetic anisotropy energy. Considerable effort has focused on molecular magnets,[32] as they offer additional applications in the areas of quantum computation and molecular spintronics.[33-35] The functionality of the devices envisioned with these technologies depends sensitively on the interplay between the exchange coupling of the magnetic cores in the dimer and the degree of delocalization involved in the electron transfer between cores.[36]

Clearly, the development of a local moment and ferromagnetism is correlated to the doping level, and suggests a non-zero spin must develop on the Fe atoms, since none of the other elements in the material are magnetic. The importance of the Fe-dimers to the magnetic behavior observed is evident in Fig. 3b, which indicates that $\mu_{\text{eff}}$ / Fe is very small. The magnetic and thermal data are consistent with a model where the extrinsic electrons added upon Ge doping result in a one-electron reduction of the Fe dimers. The lowest energy process to create a mixed-valence dimer by reduction [Fe(III)-Fe(II)], is one in which the Fe moments in the dimer remain anti-aligned with a net spin of $\Delta S = \frac{1}{2}$.[36] Figure 7 represents a cartoon picture of the proposed model where we assume, as was speculated by Yin and Pickett,[10] that the non-magnetic ground state in pure FeGa$_3$ is due to Fe(III) atoms forming a collection of singlet dimers. In this state, each dimer has a net spin of zero (Fig. 7a). The Fe-Fe distance in the dimer is largely insensitive to Ge doping at the concentrations we studied, thus we expect the dimer structure to survive. Each Ge atom that replaces a Ga also adds one electron, which then participates in the reduction of a dimer, resulting in a net spin of ½ (Fig. 7b). Clearly these electrons are fairly delocalized since: (i) there is an insulator-to-metal transition and metallic conductivity for even the lowest doping levels measured (<1%), (ii) the temperature-independent Pauli paramagnetic background term in the magnetic susceptibility becomes positive and grows with $x$, and (iii) the value of the Rhodes-Wohlfarth ratio (Fig. 3c) is considerably larger than 1.

The magnetic structure in the doped series can then be considered as a collection of interacting, mixed-valence, net-spin-½ Fe dimers. The small size of the effective moment is consistent with only a fraction of the dimers carrying this net spin. Based on one formula unit, there are 12 Ga atoms and 4 Fe atoms (2 dimers) per unit cell. Since each Ge atom adds one electron, there will be $12x$ electrons per unit cell, or $6x$ electrons per dimer. In the model, each extra electron creates a spin ½ on each dimer. Thus, the effective spin associated with each dimer will be $3x$. The effective magnetic moment per dimer in Bohr magnetons as a function of $x$ is then calculated by: $\frac{\mu_{eff}}{\mathbf{dimer}} = 2\sqrt{S_D(S_D+1)} = 2\sqrt{3x(3x+1)}$, where $S_D$ is the effective spin per dimer as a function of $x$. The solid line in Fig.3b is a fit to $\mu_{eff}/\text{Fe}$ (i.e. the dimer effective moment divided by 2) based on this simple counting picture, where the extrinsic electrons are uniformly distributed over the dimers. In the paramagnetic state below the QCP, the model predicts the effective moment remarkably well. The effective moment should scale with the carrier density, and the sharp increase in $\mu_{eff}/\text{Fe}$ at the QCP coincides with a similar feature in the carrier density, $n_H$, as measured by the Hall effect (Fig. 3b, right axis). The inset of Fig. 3b shows that the effective moment per $x$ in the paramagnetic region below the QCP is nearly constant at ~1.7 $\mu_B$, consistent with a spin-½ object.

By direct integration of the $\Delta C/T$ curves (Fig. 5a), we have calculated the low temperature ($T$ < 10 K) magnetic entropy per $x$ (Fig. 5c). The entropy is observed to peak at the QCP, which is expected, as the system tries to drive more entropy toward $T$ = 0 K at $x = x_c$. Again, assuming a simple picture of a magnetic dimer of spin ½, we recover a full $Rln(2\bar{S}+1)$ of entropy by 10 K for the sample with $x = x_c$. Interestingly, in the ordered state, for $x > 0.052$, FL behavior returns, and the magnetic entropy recovered by 10 K decreases. Presumably one would have to integrate up to and through the Curie temperature in the ordered state to fully recover $Rln2$.

## IV. CONCLUSION

In summary, we have observed a clear, and rare, FM-QCP in single crystals of the diamagnetic insulator FeGa$_3$ when doped with Ge. NFL behavior appears in the magnetic, transport, and thermal properties at a critical doping level of 5.2%. A simple empirical model where we consider a one-electron reduction of the dimer singlets, thereby creating a mixed valence state with effective spin ½, captures the main features of the physical data. The existence of an inter-dimer exchange interaction in FeGa$_3$ appears to be an essential requirement for the development of the long-range magnetic order. The density of these dimers plays the role of a novel tuning parameter for the system's quantum critical behavior. Finally, we speculate that the absence of magnetic order[16] in the series Fe$_{1-y}$Co$_y$Ga$_3$ may result from subtle changes in the structural symmetry of the FeGa$_3$ and CoGa$_3$ structure types. FeGa$_3$ forms in a centrosymmetric structure ($P4_2/mnm$), while CoGa$_3$ adopts the noncentrosymmetric structure type – ($P\bar{4}n2$). In the past, there has even been disagreement over the structure type of these two materials, as they are difficult to distinguish with powder X-ray diffraction.[19] Given the sensitivity of the material's physical properties to chemical strain, it is reasonable to expect that even small changes in the centrosymmetry could disrupt the dimer exchange and preclude magnetic order. In any case FeGa$_3$ provides a model system where FM order mediated by an inter-dimer exchange interaction near a QCP can be studied.

## ACKNOWLEDGEMENTS

DPY acknowledges many insightful discussions with Dana Browne and support from the NSF under Grant No. DMR1005764.

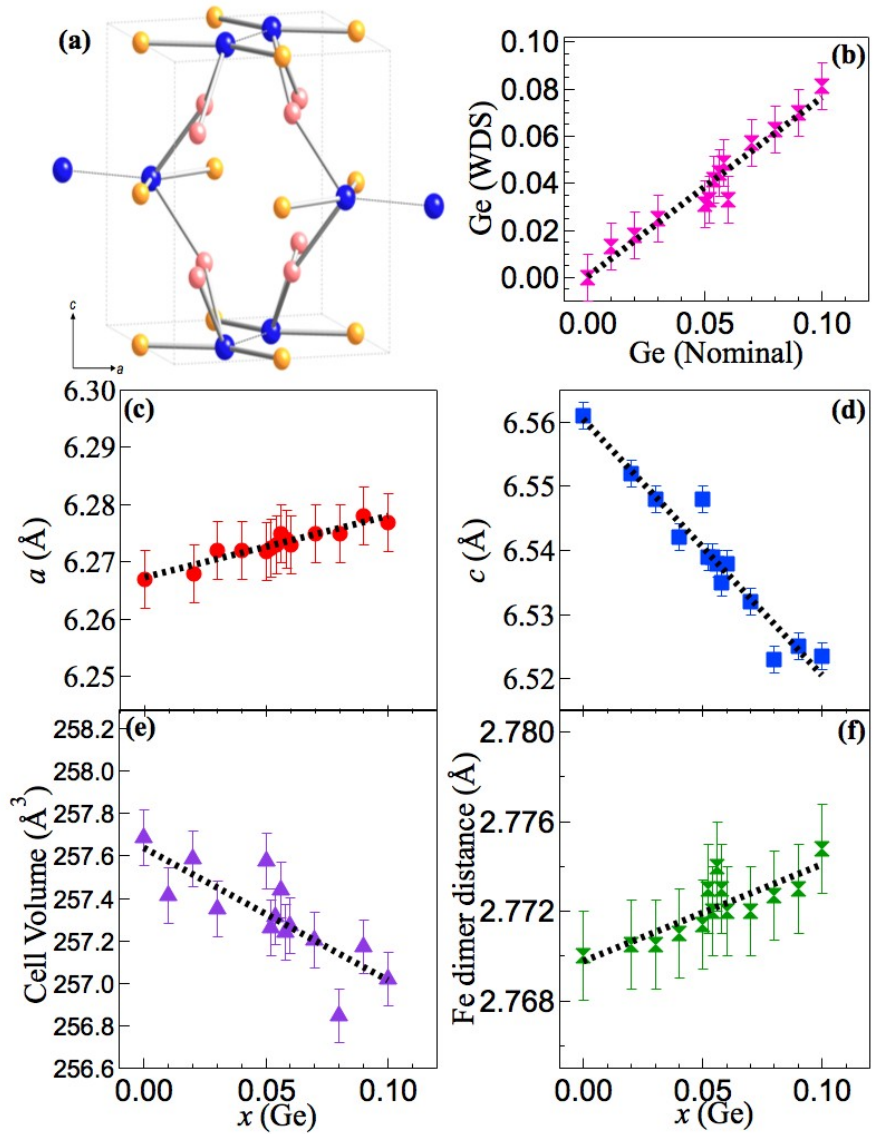

FIG. 1. (a) Tetragonal crystal structure of FeGa$_3$. The location of the Fe and Ga atoms is indicated. Fe-Fe dimers exist along the (110) direction in the $z = 0$ plane and along the $(1\bar{1}0)$ direction in the $z = ½$ plane. (b) Concentration of Ge in doped samples as measured by WDS. (c), (d), and (e) Variation of the $a$-axis and $c$-axis lattice parameters, as well as the unit cell volume as a function of doping, respectively. (f) The Fe-Fe dimer distance as function of doping.

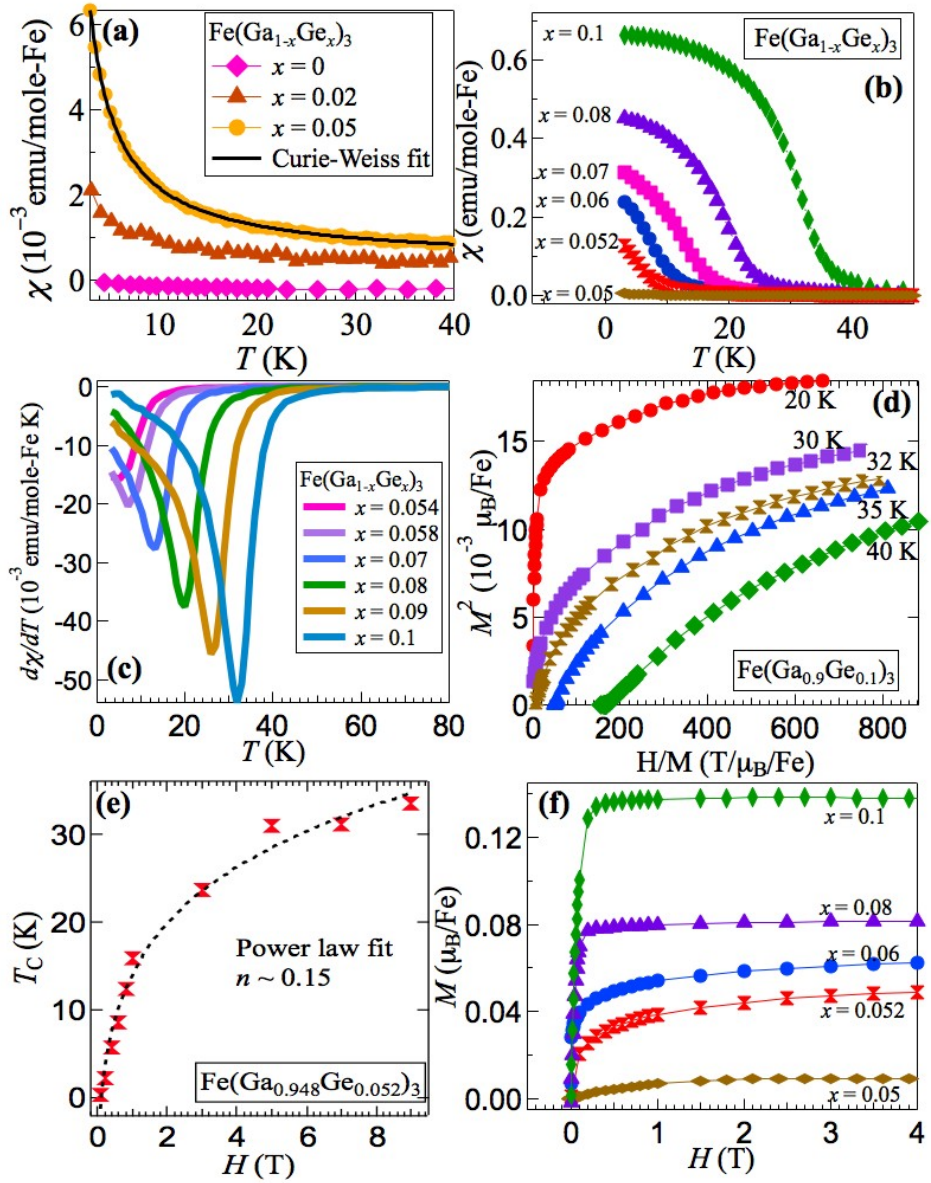

FIG. 2. (a) and (b) Magnetic susceptibility versus temperature of single crystals of Ge-doped FeGa$_3$. The solid line in (a) is a Curie-Weiss law fit to the data as described in the text. (c) Temperature derivative of the curves in (b), clearly showing the midpoints of the FM transitions. (d) Arrott plot for the $x = 0.10$ sample. The isotherm passing through zero is 32 K. (e) The magnetic field dependence of $T_c$ is shown for the $x = 0.052$ sample. (f) Magnetization versus applied field for different values of $x$.

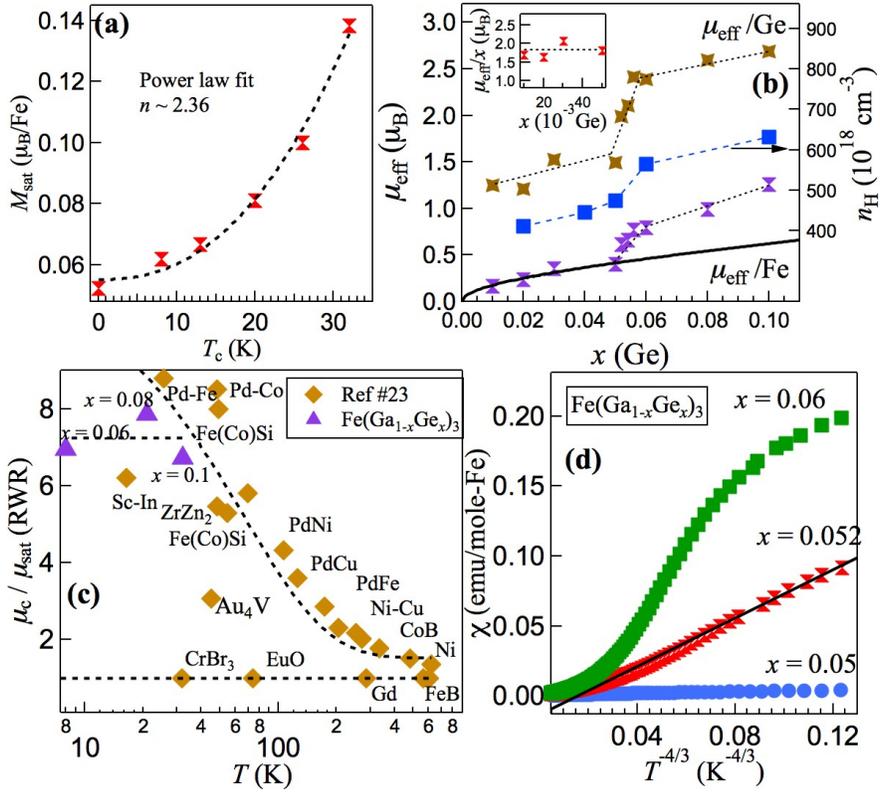

FIG. 3. (a) The saturation magnetization scaling with Tc in the series Fe(Ga$_{1-x}$Ge$_x$)$_3$. (b) $m_{eff}$ versus $x$. The moment per Fe atom increases with increasing $x$, and the solid black line is a fit to the effective moment based on the dimer model as described in the text. The inset shows the effective moment per $x$ is nearly constant at ~1.7 $m_B$, consistent with a spin ½ object. Right axis: Hall effect data showing a sharp increase in the effective moment near $x = 0.052$ coincides with a similar feature in the carrier density ($n_H$). (c) Rhodes-Wolfarth plot for various FM materials showing the itinerant nature of Ge-doped FeGa$_3$. $\mu_c = \sqrt{1 + \mu_{eff}} - 1$, and $\mu_{sat}$ is the saturation magnetization. (d) $c$ versus $T^{4/3}$ showing a linear dependence at the critical concentration $x = 0.052$.

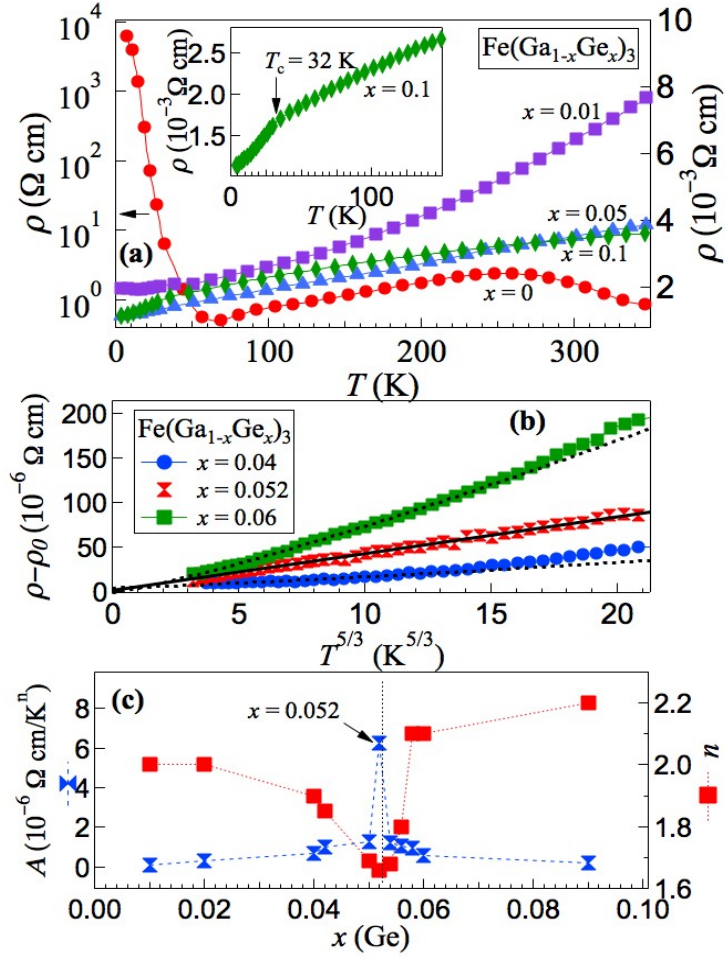

FIG. 4. (a) Electrical resistivity versus temperature of pure $FeGa_3$ (left axis) and Ge-doped samples (right axis). The inset shows the FM sample with $x = 0.10$ displays a feature in its resistivity where a decrease in the spin-disorder scattering occurs at the Curie temperature. (b) Resistivity versus $T^{5/3}$ for $x = 0.04$, $0.052$, and $0.06$. Only at the critical concentration $x = 0.052$ is a linear dependence (solid line) observed. The PM sample ($x = 0.04$) and the FM sample ($x = 0.06$) follow a $T^2$ dependence (dashed lines), consistent with Fermi-liquid behavior, and or electron-magnon scattering in the ordered state. (c) The generalized Fermi coefficient ($A$) and temperature exponent ($n$) as a function of Ge doping. The parameters were calculated by fitting the low-temperature resistivity data to the power law: $\rho(T) = \rho_o + AT^n$. Notice that $A$ is maximum at $x = 0.052$, and $n$ attains a value near $1.67$, consistent with the behavior expected near a FM-QCP.

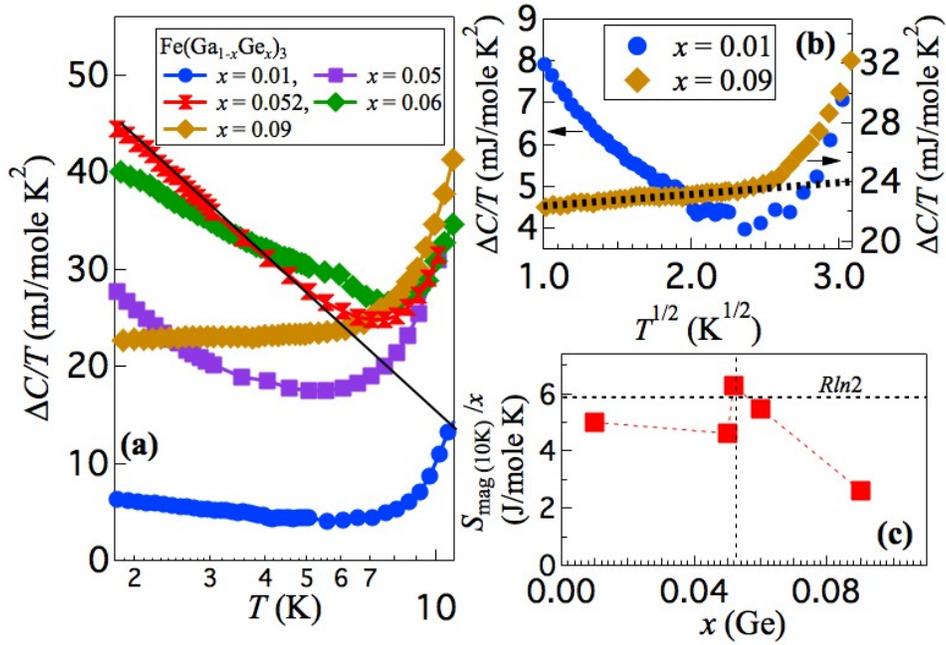

FIG. 5. (a) Electronic specific heat capacity versus temperature on a semilog plot for $x = 0.01$, 0.05, 0.052, 0.06, and 0.09. The low-temperature specific heat $\Delta C/T \sim -\ln T$ at the critical concentration $x_c = 0.052$. (b) Low-temperature specific heat showing the difference between a sample in the disordered state ($x = 0.01$) and a FM sample in the ordered state ($x = 0.09$). The dashed line shows the temperature dependence predicted for a magnon contribution to the specific heat. (c) Magnetic entropy at 10 K per $x$ calculated by direct integration of the curves in (a). The maximum entropy is observed at the critical point with a value near $R\ln 2$, consistent with a spin ½ system.

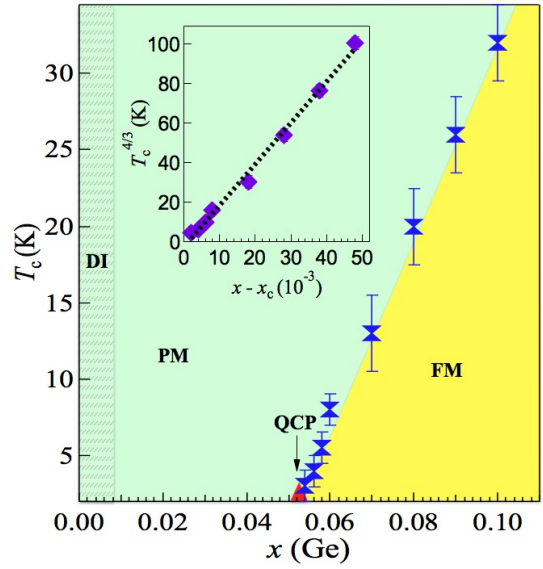

FIG. 6. Proposed phase diagram for Fe(Ga$_{1-x}$Ge$_x$)$_3$ showing the diamagnetic insulator (DI), paramagnetic (PM), and ferromagnetic (FM) phases as a function of $x$. A QCP is observed at $x = 0.052$. Inset: $T_c^{4/3}$ versus $(x - x_c)$ is linear, as expected near a FM-QCP.

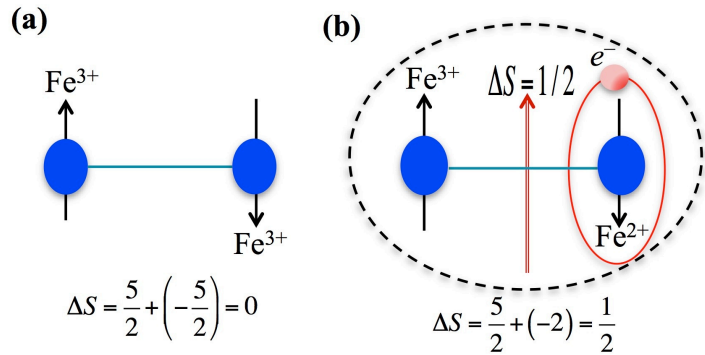

FIG. 7. Cartoon picture illustrating the one-electron reduction of a non-magnetic singlet (a) into a mixed-valence dimer (b) with a net spin of ½.

15